\documentclass[12pt]{article} 
\usepackage[T1]{fontenc}
\usepackage{amsmath}
\usepackage{amssymb}
\usepackage{amsthm}
\usepackage{amsfonts}
\usepackage{nccmath}
\usepackage{multicol}
\usepackage{geometry}
\usepackage{graphicx}
\usepackage{epstopdf}
\usepackage{epsfig}
\usepackage{wrapfig}
\usepackage{url}
\usepackage{color}
\usepackage[labelfont=bf]{caption}
\usepackage{soul}
\usepackage{floatrow}
\usepackage{authblk}
\usepackage{lineno}

\title{A stochastic model of ant trail following \\ with two pheromones}
\author[1]{Miriam Mal\'{i}\v{c}kov\'a}
\author[2]{Christian Yates}
\author[3,4]{Katar\'{i}na Bo\v{d}ov\'a}

\affil[1]{Slovak Academy of Sciences, Department of Mathematics, Bratislava, Slovakia}
\affil[2]{Department of Mathematical Sciences, University of Bath, Claverton Down, Bath, BA2 7AY}
\affil[3]{Institute of Science and Technology Austria (IST Austria), Am Campus 1, Klosterneuburg A-3400, Austria}
\affil[4]{Department of Applied Mathematics and Statistics, Faculty of Mathematics, Physics and Informatics, Comenius University, Mlynsk\'a Dolina, 842\,48 Bratislava, Slovakia}

\date{}
\begin{document}
\maketitle

\begin{abstract}
\normalsize
Colonies of ants are systems of interacting living organisms in which interactions between individuals and their environment can produce a reliable performance of a complex tasks without the need for centralised control. Particularly remarkable is the process of formation of refined paths between the nest and food sources that is essential for successful foraging. We have designed a simple stochastic off-lattice model of ant foraging in the absence of direct communication. The motion of ants is governed by two components - a random change in direction of motion that improves ability to explore the environment (facilitating food discovery), and a non-random global indirect interaction component based on pheromone signalling. Using numerical simulations we have studied the model behaviour in different parameter regimes and tested the ability of our model ants to adapt to changes in the external environment. The simulated behaviour of ants in the model recapitulated the experimentally observed behaviours of 
real ants.  Furthermore we observe interesting synchronization in ant movement patterns which vary with the diffusion properties of the pheromones.
\end{abstract}

\section{ Introduction} 

Ant colonies are highly organized complex living systems that are able to solve complex tasks at the group level without centralised control. In particular we are interested in foraging behaviours in which ants are able to form and follow refined pheromone trails between their nest and a food source \cite{four}. 

The study of collective animal behaviour, in particular the behaviour of ants, can be undertaken from a range of different perspectives. In particular one can study  generic multi-agent complex systems in which interactions between a large number of agents, combined with noise, gives rise to global patterns. 
Specifically one can characterize systems of ants directly to characterise patterns in ant colonies \cite{sumpter}. Findings in the field of collective behaviour can often be applied to different fields in order to solve seemingly unrelated problems. Such a  directed strategy of drawing inspiration from processes in nature is known as biomimicry \cite{ratnieks}. A well known use of ant models is in solving optimization problems using so called `ant colony optimization algorithms' (ACO) \cite{dorigo,doerr}.

The anatomy and the abilities of individual ant differ between (and even within species). For example, some ants use sight or direct interactions between individuals for accurate navigation, but for many species communication is primarily based on deposition and detection of pheromones \cite{jackson2,steck}. These pheromones differ in their purpose and in their physical properties \cite{duss3,ever}, which are the subject of ongoing study \cite{morgan}. Most of the models of ant behaviour therefore assume pheromone-based decision making in the movement of individual ants \cite{couzin,panait}.

The rules of communication and behaviour of ants are becoming increasingly well understood but vary between different species. Mathematical modelling is a tool which can distinguish between different biological hypothesis relating to ant behaviour \cite{bicak,edelstein,erment,watm}. 
The behaviour of a colony, driven by both local and global interactions, can demonstrate a rich variety of emergent behaviours including synchronization, oscillation, milling and switching to name but a few. These patterns are not necessarily ant-specific but are a consequence of interactions in complex multi-agent systems, often in the presence of noise \cite{detrain,dene2,sole,boi,bona,yates}. 
  
In line with the work of Couzin \cite{couzin} and Calenbuhr \cite{calen} we introduce a stochastic model of ant foraging which incorporates a minimal number of biological assumptions. Ants move by a correlated random walk \cite{kare} in an off-lattice environment on a continuous two dimensional space. The off-lattice nature of the model avoids to possibility of lattice-induced artefacts in the resulting behaviour. Communication between ants is based on two distinct pheromones that diffuse and evaporate in the environment, allowing information to both spread and dissipate. As in nature, the motion of our model ants is not deterministic. Indeed, we deliberately incorporate an element of randomness in the ants' movement rules which we will demonstrate to be an adaptive advantage in the dynamical environment \cite{dene,duss2}.
 
In what follows we give details about the explicit implementation of the model and investigate the ability of our model ants to solve basic tasks essential for colony survival. Sample challenges we pose our colony of ants are those of food source discovery and reliable trail formation between the nest and the food source even in a dynamically changing environment. We use numerical simulations to show that our proposed model can mimic realistic ant behaviour. The basic rules we impose lead to a range of emergent behaviours (such as milling or oscillatory motion of individuals) that are comparable to phenomena observed in nature \cite{dene2,couzin}. Moreover, under certain circumstances the motion of model ants can demonstrate an unexpected synchrony in nest or food discovery. This degree to which the ants synchronise depends upon the diffusion properties of the pheromones.
 
\section{Mathematical model of ant foraging with two pheromones}

We propose a model of a single colony of ants, based in its nest and study the formation of a trail to food located in a close vicinity of the nest. The model  is based on the following assumptions:
\begin{enumerate}

\item Ants do not interact directly, only through pheromones.

\item Ants move at a constant speed, the same for all ants.

\item The motion of each ant is governed by a correlated random walk.

\item Ants use two different attractive pheromones \cite{jackson2007}; one for nest-orientation \cite{steck} and one for food discovery \cite{duss3,morgan}.

\item Pheromones diffuse and decay \cite{robinson}.

\item The amount of pheromone that an ant may carry is limited.

\item Ants alter their motion in response to the pheromone concentration which they sense using their antennae and process using a Weber's law \cite{calen},  \cite{perna}.

\item Ants switch between two phases: food-searching and nest-searching.

\item There are no births or deaths on the time-scale of interest (i.e. the total number of ants is constant).

\item Ants move on an infinite two dimensional plane.
\end{enumerate}

The model can be characterised by four distinct phenomena: the random motion of ants, the deposition of pheromones by ants, the physical process of pheromone diffusion and decay in the environment and the response of ants to the perceived pheromone signal.

\subsection{Motion of ants}
  
\begin{figure}[t]
	\includegraphics[width=13cm]{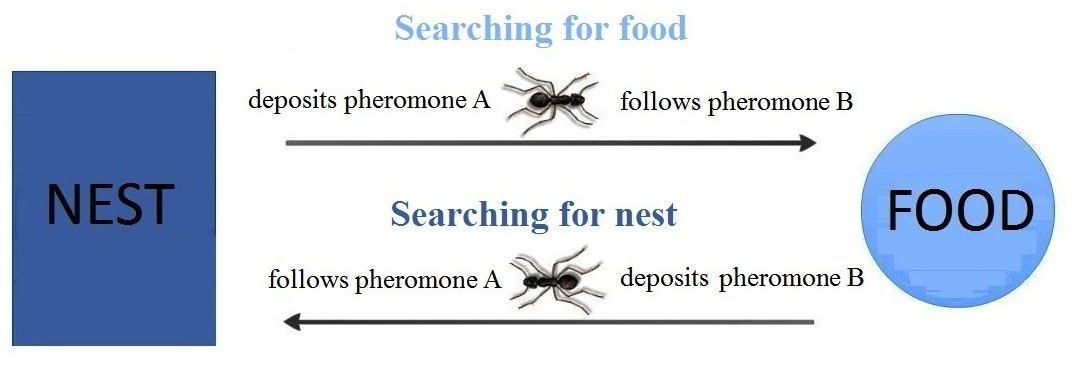}
\caption{A model of foraging ants with two task-dependent pheromones. 
Communication between the ants in the model is mediated through changes to the environment by deposition of two attractive pheromones. These pheromones have different roles, as indicated in the figure. When ants are searching for food they deposit pheromone A and follow pheromone B. Those searching for the nest deposit B and follow A. The two pheromones may vary in their physical properties (diffusion coefficient and rate of decay for example) but also in the amount deposited by ants over time or their ease of detection. 
\label{diagram}}
\end{figure}

\begin{figure}[t]
\includegraphics[width=5cm]{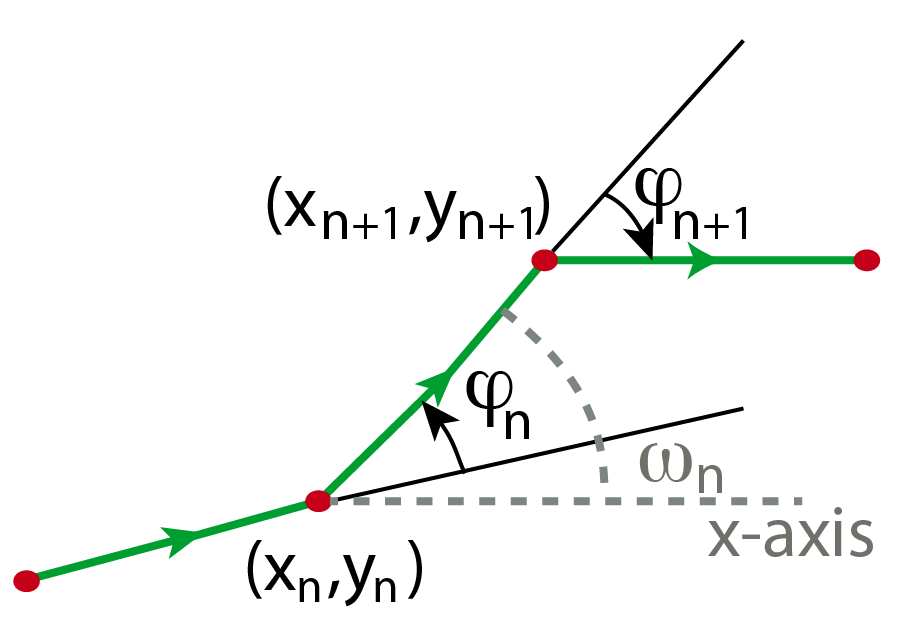}
\caption{The motion of a single ant consists of a series of discrete steps in space. In our fixed time-stepping algorithm the constant speed of the ants corresponds to a fixed spatial step size, $L$. The direction of motion at a given step is denoted by $\omega_n$ while the difference between two consecutive step directions, referred to as the angle deviation, is denoted by $\varphi_n$. The angular deviation comprises two components - a deterministic term $f(c)$ which depends on the concentration of detected pheromone and a stochastic term $\sigma(c) \xi_n$ that accounts for imprecision of motion and also depends on the pheromone concentration. The heading $\omega_n$ is ant's original direction plus the sum of all directional changes until the $n$-th step. 
\label{fig_direction}}
\end{figure}

We model the motion of the ants using a correlated random walk \cite{codl1,codl2,codl3} specified as follows:
\begin{align} \label{CRW}
\omega_{n+1}^i &= \omega_n^i + f(c) + \sigma(c) \xi_n^i
\end{align}
where $\omega_n^i$ represents the direction of motion in the $n$-th step of the $i$-th ant (see figure~\ref{fig_direction}). The angular deviation between steps for ant $i$ is given by $\varphi_n^i = f(c) + \sigma(c) \xi_n^i$. The first term, $f(c)$, represents the deterministic response of an ant to the perceived pheromone concentration, $c$, in its direction of motion and the second term, $\sigma(c) \xi_n^i$, is an angular noise whose magnitude depends on its perceived pheromone concentration. 

In the absence of pheromones the directional change for the $i$-th ant in the $n$-th step is $\varphi_n^i=\sigma \xi_n^i$. This represents a normally distributed random variable with a zero mean and constant variance, $\sigma^2$, when $\xi_n^i$ is drawn from the standard normal distribution. 

The randomness in the movement of individual ants  is important for the behaviour of the whole colony. We will demonstrate that this noise at the individual-level plays a significant role in each of the phases of colony behaviour - foraging, trail formation and adaptation to changes in the environment. 

The position of an ant $i$ in Cartesian coordinates is calculated according to the following update rules
\begin{eqnarray}
x_{n+1}^i &=& x_n^i+ L\cos (\omega_{n+1}^i) \label{X},\\
y_{n+1}^i &=& y_n^i+ L\sin (\omega_{n+1}^i)\label{Y},
\end{eqnarray}
where $L$ is the constant spatial step size. Bicak \cite{bicak} suggests that, given time steps of 0.5 seconds, this step size, $L$, corresponds to a distance of two ant body lengths. Bicak also demonstrated experimentally for {\it Pharaoh's} ants that the angular deviations between discrete, fixed time steps can be well approximated by a normally distributed random variable. 

\subsection{Pheromone deposition}

The ants can deploy two pheromones that differ in their purpose and in their physical properties. Pheromone A is employed for nest location, whilst pheromone B signals the location of a food source. Ants are able to distinguish between the two pheromones and to determine their intensity.

We assume that all ants in the system are in one of the two states - either they are foraging (i.e. searching for food) or they are searching for the nest. Each ant deposits and follows a pheromone according to its state. Deposition of pheromones serves to provide information about the system and the environment to other ants. An ant deposits a unit of a pheromone A at each step (whilst its pheromone is not depleted) when it is in the food-searching state, and a unit of pheromone B when it is in the nest-searching state. An ant follows pheromone B when in the food-searching state and pheromone A when in the nest-searching state as shown schematically in figure~\ref{diagram}. 

Ants deposit pheromones only for a fixed number of steps, starting when entering the respective state. This represents a limited supply of the pheromone for each ant in one excursion, replenished when an ant returns to the nest. For reference, all of the parameters of the model are given in table~\ref{tab1}.

\subsection{Diffusion and decay of pheromones}

Pheromones deposited in the environment diffuse and decay/evaporate in time. Function $c(x,y,t)$ gives the concentration of a given pheromone at location $(x,y) \in \mathbb{R}^2$ at time $t$. The function $c(x,y,t)$ evolves according to the two dimensional diffusion equation: 
\begin{equation} \label{dif}
\frac{\partial c}{\partial t}= D \left ( \frac {\partial ^2 c}{\partial x^2} + \frac {\partial ^2 c}{\partial y^2} \right ) \,,
\end{equation}
where $D$ is a diffusion constant  for given pheromone. In our model we consider two types of pheromones. Therefore, the description of the system requires two independent diffusion equations for $c_A(x,y,t)$ and $c_B(x,y,t)$ with respective diffusion coefficients $D_A$ and $D_B$ (with values given in table \ref{tab1}).

Since the positions and headings of all ants are updated synchronously, each ant deposits pheromones a fixed amount of pheromone at their current location simultaneously in every step. These pheromone depositions can be represented as additional point-source terms making the problem time inhomogeneous. However, due to the linearity of the diffusion equation, we can decouple the problem into many independent subproblems corresponding to each deposition event and solve the full problem using a superposition of the individual solutions. The concentration of each pheromone deposit evolves according to the diffusion equation \eqref{dif} (with the appropriate diffusion coefficient) and has the following initial condition
\begin{equation} \label{IC}
c_{ij}^0(\boldsymbol{x}|\boldsymbol{x}_{i,j},t_{ij})=m\delta(\boldsymbol{x}-\boldsymbol{x}_{i,j}),
\end{equation}
where $m$ is the amount of pheromone deposited by ant $i$ at {the $j$-th} step, {$\delta(\boldsymbol{x}-\boldsymbol{x}_{i,j})$} is a Dirac delta function and  {$\boldsymbol{x}_{i,j}=(x_{i,j},y_{i,j}) \in \mathbb{R}^2$} is the position of the depositing ant, $i$, at the time of deposition time, $t_{ij}$. Here $j$ indexes the times at which ant $i$ deposits pheromone.
The solution for one of these subproblems at time $t$ is given by the fundamental solution to the diffusion equation: 
\begin{equation}
c_{i,j}(x,y,t)= \frac{1}{4\pi D (t-t_{i,j})}\exp\left[ -\frac{(x-x_{i,j})^2 + (y-y_{i,j})^2}{4\pi D (t-t_{ij})}\right],
\end{equation}
for $t>t_{ij}$.
The exact solution of the full time-inhomogeneous problem is therefore a sum over the independent solutions corresponding to the each of the deposition events from each of the ants that have occurred up to that point in time
\begin{equation} 
\label{solution}
c(x,y,t)=\sum_{i=1}^N\sum_{j=1}^M c_{i,j}(x,y,t)\mathbb{I}_{t_{ij}<t}\,
\end{equation} 
where $M$ is the total number of pheromone depositions each ant can make (the same for each ant) and $\mathbb{I}_{t_{ij}<t}$ is the indicator function which takes the value $1$ if $t_{ij}<t$ and zero otherwise.
 
Pheromones in the system also evaporate and thus the total amount of pheromone deposited is not conserved in time. We model this by discarding contributions to the full solution from those depositions that occurred earlier than a predefined threshold time. Effectively we can represent this in the full solution by changing the range of validity of each contribution by replacing the indicator function in equation \eqref{solution} with $\mathbb{I}_{t-\delta < t_{ij}<t}$, where $\delta$ is the `pheromone lifetime' (see table \ref{tab1}).

\subsection{Response to pheromone concentration}

The pheromone concentration affects both the deterministic component, $f(c)$, and the amplitude of the stochastic component, $\sigma(c)$, in the angular update equation \eqref{CRW}. We assume that the deterministic response function depends solely on concentrations at the tips of the antennae i.e. $f(c) = f(c_{-\frac{\pi}{6}},c_\frac{\pi}{6})$. Here  $c_\frac{\pi}{6}=c(x_i+l\cos(\omega_n^i+\pi/6),y_i+l\sin(\omega_n^i+\pi/6))$ is the pheromone concentration at the left antenna of ant $i$ and $c_\frac{-\pi}{6}$ is defined analogously for the right antenna, where $l$ is the antenna length, uniform across all ants. There is evidence that organisms, including ants, are able to evaluate environmental signals and modify the direction of motion in response to them. In animals \cite{calen} a common form of the response function is known as Weber's law \cite{weber}, which been confirmed  experimentally for ants \cite{perna}. Generally ants follow the direction of higher pheromone concentration \cite{
suckling}.
The function $f(c)$ encodes the Weber's law describing the response of the $i$-th ant to pheromone concentration, $c$: 
\begin{equation}
\label{Weber}
f(c_{-\frac{\pi}{6}},c_\frac{\pi}{6})=\begin{cases}
-\frac{\pi}{6} & \text{if } \Gamma>\alpha, \text{ and }\, c_{-\frac{\pi}{6}}>c_{\frac{\pi}{6}},  \\  
\frac{\pi}{6}& \text{if } \Gamma>\alpha, \text{ and }\, c_{\frac{\pi}{6}}>c_{-\frac{\pi}{6}},\\
0 & \text{otherwise,}
\end{cases}\,
\end{equation}
where 
\begin{equation}
 \quad \Gamma= \left|\frac{c_\frac{\pi}{6}-c_{-\frac{\pi}{6}}}{c_\frac{\pi}{6}+c_{-\frac{\pi}{6}}}\right|\,
\end{equation}
is the relative difference in concentration of pheromone at the end of each antennae and $\alpha$ is a given detection threshold, commonly called the just-noticeable difference (JND). Weber's law allows ants (and analogously other animals) to efficiently process information from the environment that may span many orders of magnitude.

The magnitude of the stochastic component of the angular update equation \eqref{CRW}, $\sigma(c)$, is also affected by the detected pheromone concentration. When the pheromone density is high (above the given threshold $c_{\min}$, see table \ref{tab1}) the amplitude of the noise term decreases but never disappears completely. The function that we propose in our model is \begin{equation}
\label{sigma}
\sigma(c)=\sigma \min  \Bigg\{ 1, \frac{c_{\min}}{\max\{c_{\frac{\pi}{6}}, c_{-\frac{\pi}{6}}\}} \Bigg \}\,,  
\end{equation}
where $\sigma$ is the variance of the noise in directional change in the absence of pheromone. 

\subsection{Implementation}

In the next section we describe a number of numerical simulations in which the behaviour of the model ants is studied in a range of different environmental conditions.
for each different environment we have performed multiple numerical simulations of the system in MATLAB\textsuperscript{\textregistered}. In each simulation we consider $N$ ants with a fixed  nest position and food source(s) placed away from the nest. At the beginning of each simulation all $N$ ants are placed in the nest and there is no initial pheromone signal. 
 
The position of each ant at the next step is calculated using difference equations \eqref{X} and \eqref{Y}, where the direction of the $i$-th ant, $\omega_{n+1}^i$, describes a correlated random walk given by equation \eqref{CRW}.   

At the beginning of the simulation ants leave the nest in the food-searching state. They randomly explore the environment and forage for food. When an ant discovers a food source it changes its state to nest-searching mode. At each step each ant deposits a fixed amount of pheromone corresponding to its food- or nest-searching state. However, the amount of each pheromone that an ant can deposit in one exploration is limited. Upon leaving the nest or reaching food (resp.) an ant deposits the pheromone A for $t_A$ time steps or pheromone B for $t_B$ time steps (resp.) (see table \ref{tab1} for values). Each time-step in the simulation corresponds to {half} second of real-time.

Furthermore, not all the deposited pheromone remains in the system but it {decays} over time. The decay rate may vary for the different pheromones ($\delta_A, \delta_B$, see table \ref{tab1}). Unless otherwise stated, in our simulations, we assume that the pheromone marking the nest is more permanent than the food marking pheromone. Later we also study the case when the situation is reversed. The constants $\delta_A, \delta_B$ represent the time period for which we consider a given pheromone deposition, represented by an initial condition \eqref{IC}, to be part of the solution profile given by equation \eqref{solution}.

In order to simulate the system we need to calculate the pheromone concentration at every point ants might be found in order to determine inputs for the direction function \eqref{CRW}. In theory it is possible to do this using the analytical solution \eqref{solution}. However, generating the pheromone field at all the required points in space and time would be computationally challenging with an analytical approach. Instead we used the more efficient finite difference method for the solution of \eqref{dif} to generate the pheromone concentration field at the required times (see table \ref{tab1} for finite difference parameters).

\section{Results}

We tested our mathematical model of ant foraging in several different scenarios using the numerical simulation method described above. First, we considered a scenario with fixed nest and food positions. Both nest and food were circular in shape with radii 3 cm and 2.5 cm, respectively, and a separation distance of 15 cm between the centres. We chose the ant-specific parameters of the model to represent a common type of ant - a Pharaoh ant, with a body length of approximately 2 mm. Initially the environment is pheromone-free. The simulation parameters are summarised in a Table~\ref{tab1}. 

\begin{figure}[H]
	\begin{center}
	\includegraphics[width=11cm]{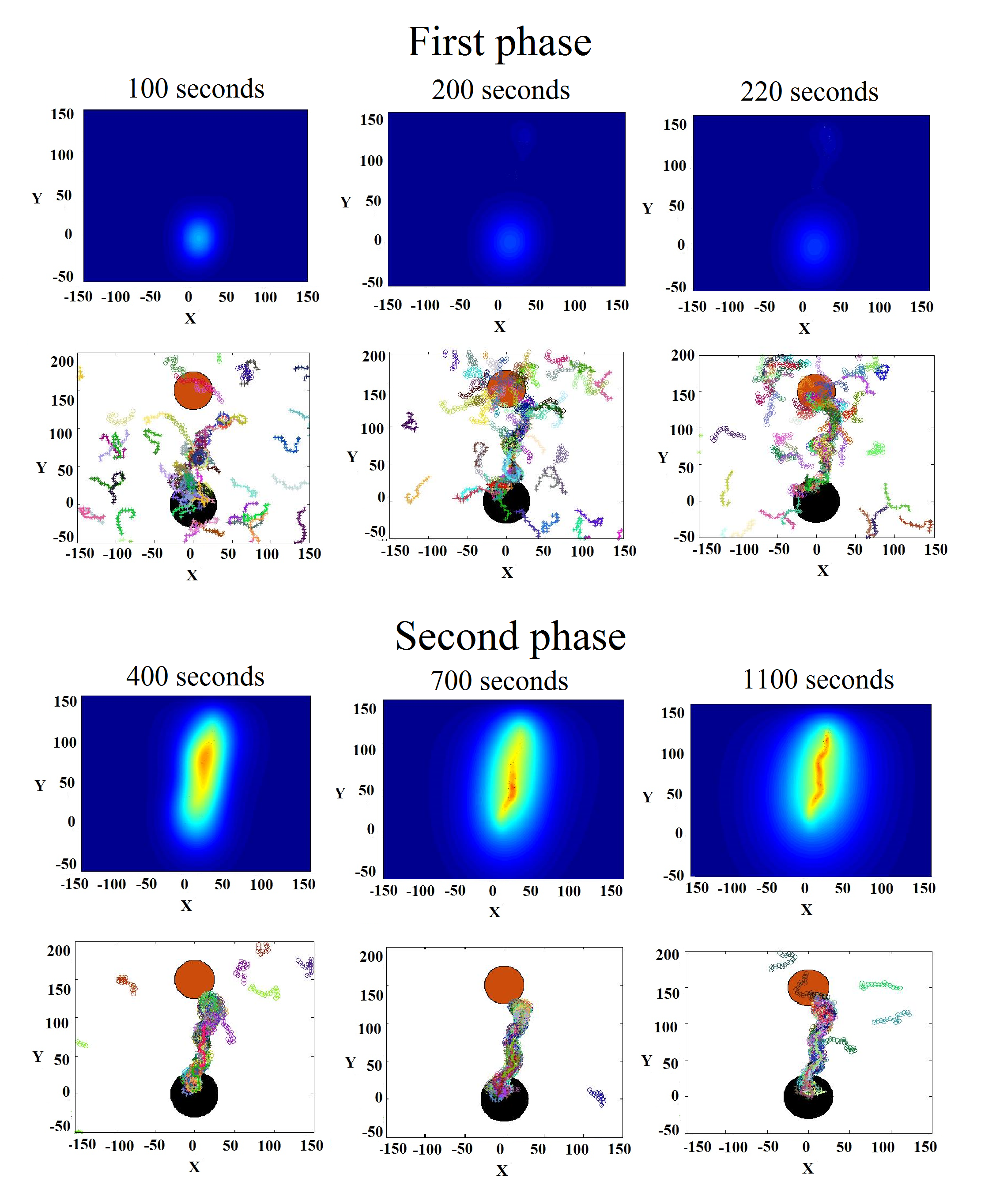}
	\caption{Formation and refinement of a trail between the nest and the food.
	The numerical simulation is divided into two phases. Initially all ants are in the nest and search for the food in the 
	environment with no pheromone signal. This first phase terminates when 10 percent of ants find the food and return back 
	to the nest; at this point the environment contains pheromone profiles marking the route to the food (pheromone A) and to the nest
	(pheromone B) and all ants are returned to the nest. In the second phase ants leaving the nest can follow the pre-existing pheromone A marking the route to the food source enabling them to refine and maintain a narrow trail between the nest and the food.
	In each set of two rows, the upper boxes show the sum of the concentrations of both pheromones in space at a given time (scale 0 - 0.0095 $g/mm^2$ from dark blue to red). Lower boxes show
	 the last 15 positions of each ant (in different colours) marking a recent trajectory. The black and red circles mark the location of the nest and the food source respectively.
}
	\label{fig3}
	\end{center}
\end{figure}

Each simulation with this set-up consisted of two phases. In the first phase, all $N$ ants started in the nest and explored the environment for food sources. At the same time they deposited pheromone A (marking the location of the nest) for a limited time. When an ant discovered the food source it changed state and started following pheromone A and depositing pheromone B. Once 10 percent of the $N$ ants had found the food source and returned back to the nest we ended the first phase of the simulation. Since the environment in the first phase of the simulation is initialised {without} pheromone, some ants may wander far from the nest (with no guarantee of returning) and effectively become lost. Such `losses' also occur in real colonies \cite{wystrach}. In order to avoid the computational expense incurred in keeping track of these ants and calculating the pheromone field over such a large area we decided to disregard these 'lost ants' in the numerical simulation by beginning a second phase of simulation. 
The second phase was initialised with the pheromone concentration field from the end of the first phase and a new set of $N$ ants in the nest. These ants follow the existing pheromones and they replenish it with the newly deposited pheromone as previously described.

Snapshots from an example simulation are presented in Figure~\ref{fig3} for the scenario with a single food source. While in the first phase, ants dispersed randomly, following the correlated random walk, after approximately 4 minutes (once 10 percent of the original $N$ ants had returned to the colony) they switched to a second phase where the pheromone trail between the food source and the nest was gradually reinforced. After approximately 8 minutes most of the ants were concentrated around a narrow pheromone trail, following it back-and-forth while switching between the states. However, there were still some ants that failed to find the trail quickly and thus explored a wider region of space before discovering the trail. We will demonstrate that this inhomogeneity in ant behaviour may be advantageous to the colony as a whole since these  `far-ranging' ants may be able to discover other sources of food or adapt to a changing environment. In reality food sources may become depleted and ants need to be able to respond efficiently to such situations. 

Since the process of trail formation is stochastic, the time to establish the trail between food and nest is a random variable. With  small probability all ants may get lost and the trail will never be established, however, the probability of such failure decays quickly when the number of the ants is increased. With $N=100$ ants we observed trail formation every simulation we ran.

In our model the adaptability of ants is facilitated by the limited persistence of pheromone gradients due to decay or diffusion. By choosing the lifetime of the food-marking pheromone to be smaller than that of the nest-marking pheromone we allow the model ants to respond to the relative transience of the food source and the relative permanence of the nesting site. Similarly, by default, the diffusion of the food-marking pheromone is chosen to be larger than the diffusion constant of the nest-marking pheromone so that information about the transient food source is transferred more quickly than information about the location of the more permanent nest site. 

\begin{figure}[H]
	\centering
	\includegraphics[width=11cm]{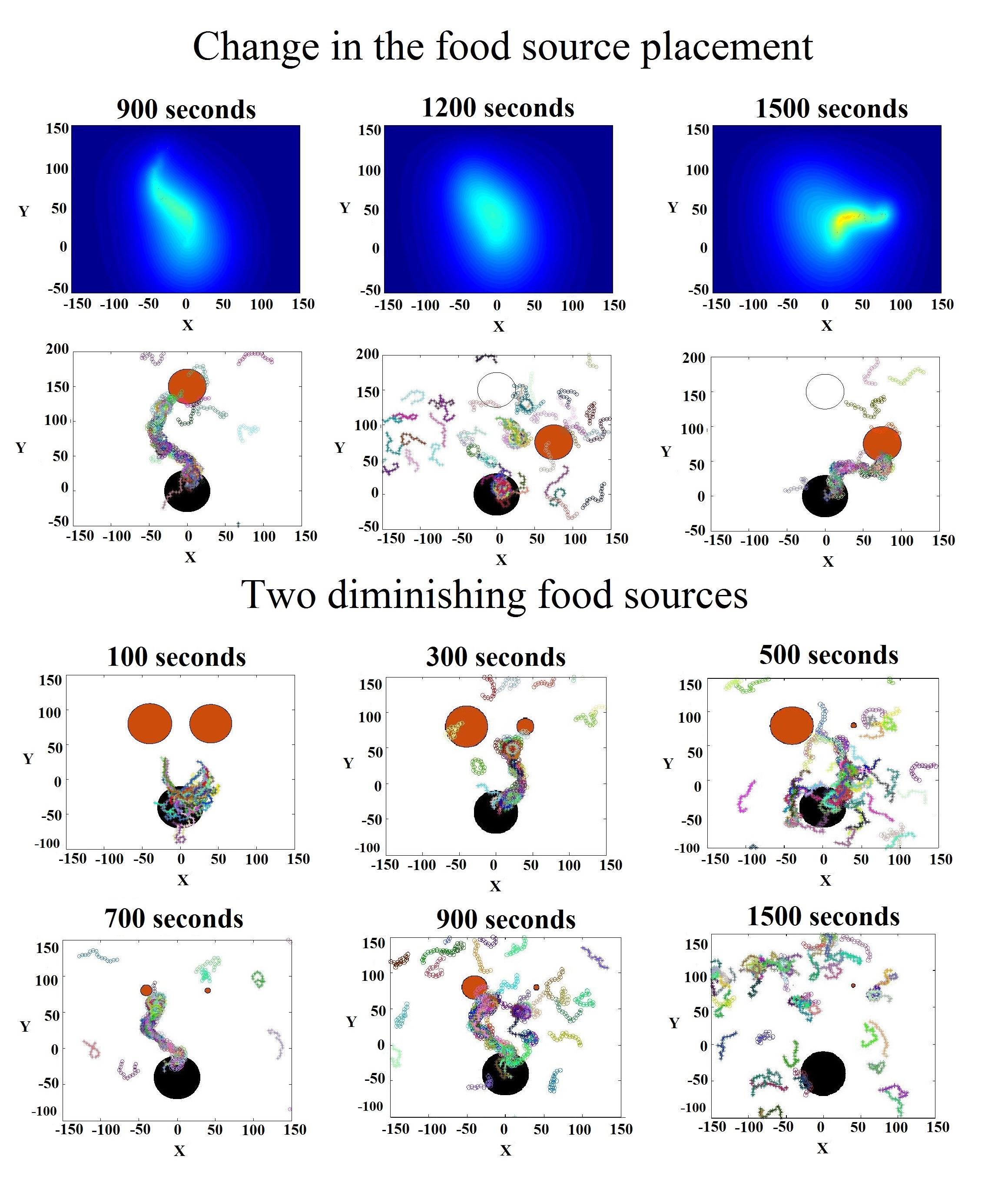}
	\caption{{Adaptation to a changing environment.} 
	We study two dynamic scenarios in which food sources change location (top two rows) or are resource limited (bottom two rows). In the first scenario (top two rows) the food location is changed after 900 seconds. Ants are able to create a new pheromone trail leading to and from the new food source (pheromone field scale is 0 - 0.0115 $g/mm^2$ from dark blue to red). 
	In the second scenario two identical food sources are present in the 
	environment but depletion of the initially-favoured food source encourages the ants to explore space and find another source of food. Figure descriptions are as in Figure~\ref{fig3}.
}
	\label{fig4}
\end{figure}
 
In order to investigate the adaptability of ant behaviour we dynamically altered the environment. In particular, we changed the location of the food source location once the trail to the original food source had been established. Finding a new food source and establishing a new trail took the model ants less than 10 minutes, and involved the decay of the original pheromone trail, random exploration of the space for other food sources and the formation of a new pheromone trail to the relocated food source. The results of an example simulation are given in the top two rows of Figure~\ref{fig4}. Within approximately 300 seconds of the change in the food source from coordinates [0,150] to [100, 100] the original pheromone trail had effectively disappeared. After an additional 300 seconds the ants had formed a new trail from the nest leading to the relocated food source. 

We also explored a situation where two food sources exist simultaneously at the same distance from the nest and are resource-limited. In order to implement food-source depletion, every time an ant discovers the food source, the amount of food available is decreased by a constant amount. Practically this is implemented by decreasing the radius of the circle which represents the area of the food, making it more difficult to find in the future. Each food source is completely depleted once visited 300 time. Simulations were initialised in the primary `exploration' phase with no pheromone in the environment. First a trail was formed to one of the (randomly chosen) food sources and, as its size diminished ants were able to find and switch to the second food source. After the second food source ran out the ants continued to explore the space for other food sources. Interestingly the first food source did not have to be completely depleted before the ants switched to the second food source, indicating that the model ants are able to discern between food sources of differing quality.  

Snapshots of an example simulation are given in the bottom two rows of Figure~\ref{fig4}. The simulation was initialized with two equidistant food sources that gradually diminished every time an ant that visited that source. Initially the ants formed a trail to a single food source. When this first food source was mostly consumed they discovered the second and formed a new trail connecting it to the nest. After both food sources we largely consumed ants started to forage in the environment at random. 
   
\begin{figure}[H]
	\includegraphics[width=14cm]{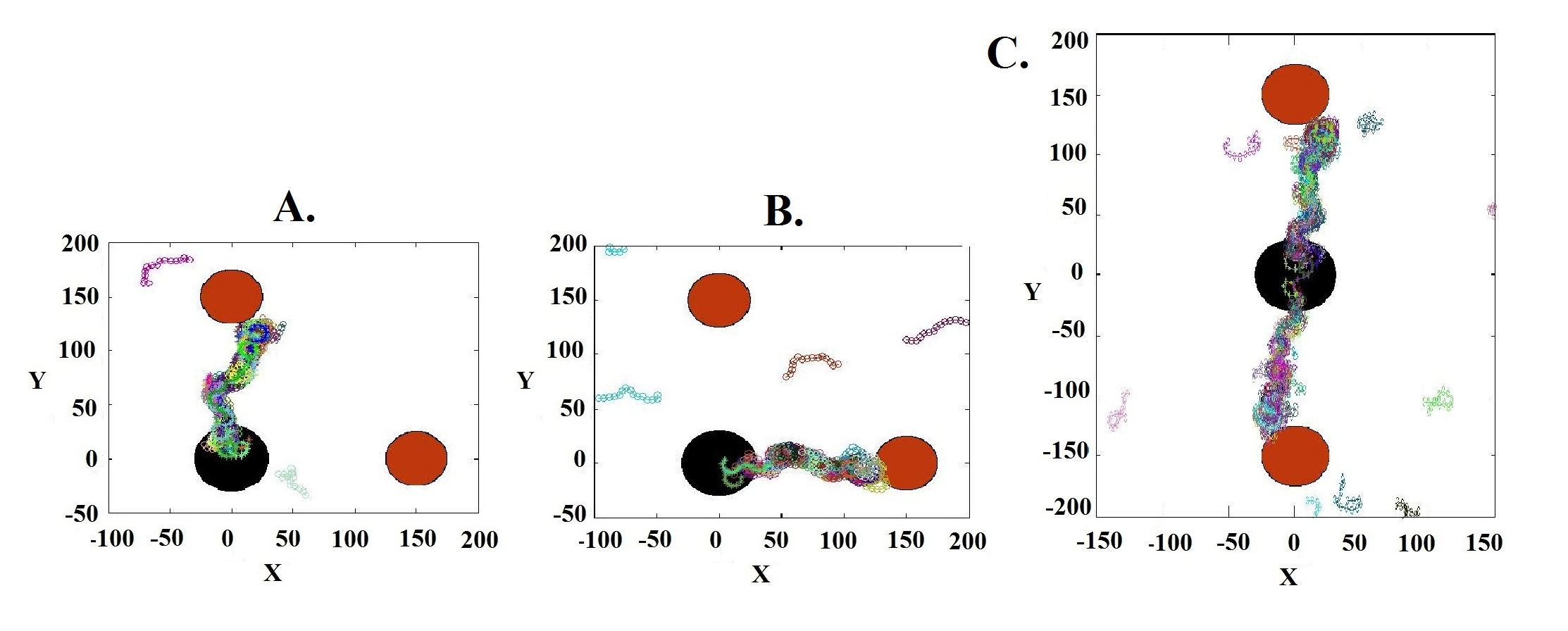}
	\caption{Two food sources were placed at equal distances from the nest at different positions and we studied: (i) whether the colony forms one or two paths, and (ii) whether ants are able to maintain both paths. We initialized the pheromone distribution using a superpositions of subsystems with only one food source.  In (C), the simulation with $N=200$ ants and the food placed on the North side was superimposed with the same pheromone concentration field reflected around the horizontal axis while in (A) and (B) the subsystem with $N=150$ ants was rotated by angle $\pi/2$. All ants were placed to the nest at the in the environment with two food sources and the described pheromone field, where each panel represents a different simulation displayed at 1200 seconds. In panels (A) and (B) a single path formed to one or other of the food sources with equal probability but almost never to both simultaneously. In panel (C) ants formed two trails leading to both food sources with a high probability.
}
	\label{fig5}
\end{figure}

We next explored whether our model ants are able to take advantage of multiple food sources and maintain multiple trails. We placed two food sources at the same distance away from the nest but at different subtending angles. The results of some example simulations with different subtending angles are given in Figure~\ref{fig5}. When the angle between the two food sources was $\frac{\pi}{2}$ ants formed a trail to either one of the two food sources with equal probability (see Figure~\ref{fig5} (A) and (B)). However, they were not able to maintain trails to the two food sources simultaneously, even when the pheromone trail to both food sources was initialised, see the legeng in Figure~\ref{fig5}. On the other hand, if the food sources were placed on diametrically opposite sides of the nest, ants were able to  maintain both trails at the same time. If the angle between the food sources is sufficiently small then the two food sources are effectively in competition with each other for trail formation. Symmetry breaking occurs because 
stochastically more ants find one of the food sources earlier than the other. The increased pheromone deposition on this trail leads to increased recruitment to the food source. This stochastic positive reinforcement of the trails leads to the choice of one of the food sources over the other with equal probability. However if the subtending angle between the two food sources is too large the two trails are essentially independent and can both establish.

We also looked at the dynamics of individual ants on an established path between the nest and the food source in order to investigate the distribution of ants within a trail. We found that the mathematical model admits at least two phenomena also observed in nature \cite{couzin}. The first one is an oscillating motion of ants within a trail. The ants are not able to follow the maximum pheromone gradient precisely due to the random component of their motion. As an ant walks in the direction of the steepest ascent the difference between the pheromone concentrations at its antennae, which contributes to the deterministic part of the motion ($f(\cdot)$), becomes small and the noise term dominates taking the ant away from the pheromone peak. As soon as the ant leaves the peak pheromone region it is attracted back again due to non-zero deterministic component. This results in an oscillating motion around the path of the maximal pheromone. 

The second biologically realistic emergent phenomenon exhibited by our model is that ants may spontaneously form circular mills \cite{dene2}. In our model these mills are a result of ants who are searching for food following the pheromone traces of the ants searching for the nest (and vice versa) in a particular spatial configuration. Since their local pheromone signals are compatible ants may follow each other in a loop. This has also been observed in nature and may have disastrous consequences for ant colonies, in particular in the case of strongly following ants, (e.g. army ants) which sometimes get trapped in a so called ``death spiral'' in which the majority of the colony forms a positively reinforcing mill \cite{delsuc}.     

\begin{figure}[H]

\includegraphics[width=0.49\textwidth]{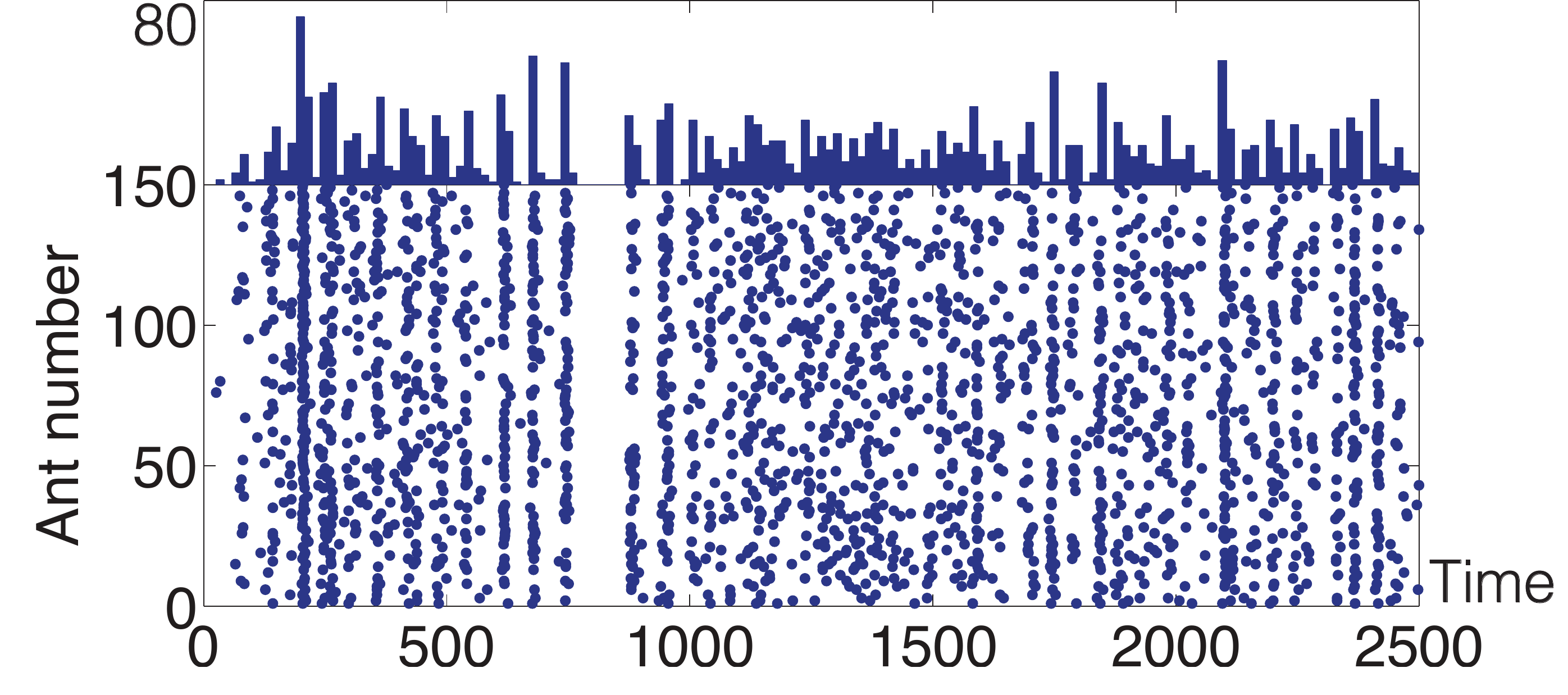}
\includegraphics[width=0.49\textwidth]{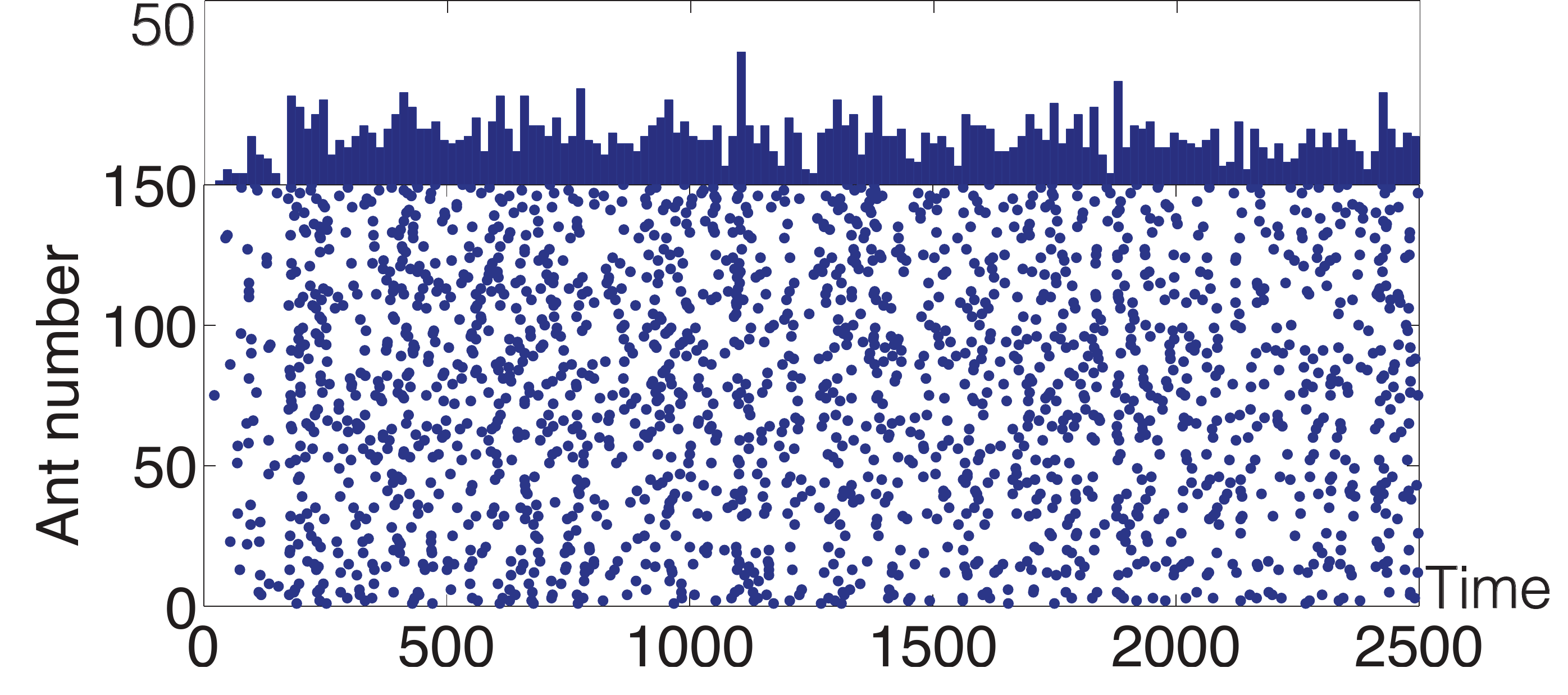}\\
\includegraphics[width=0.49\textwidth]{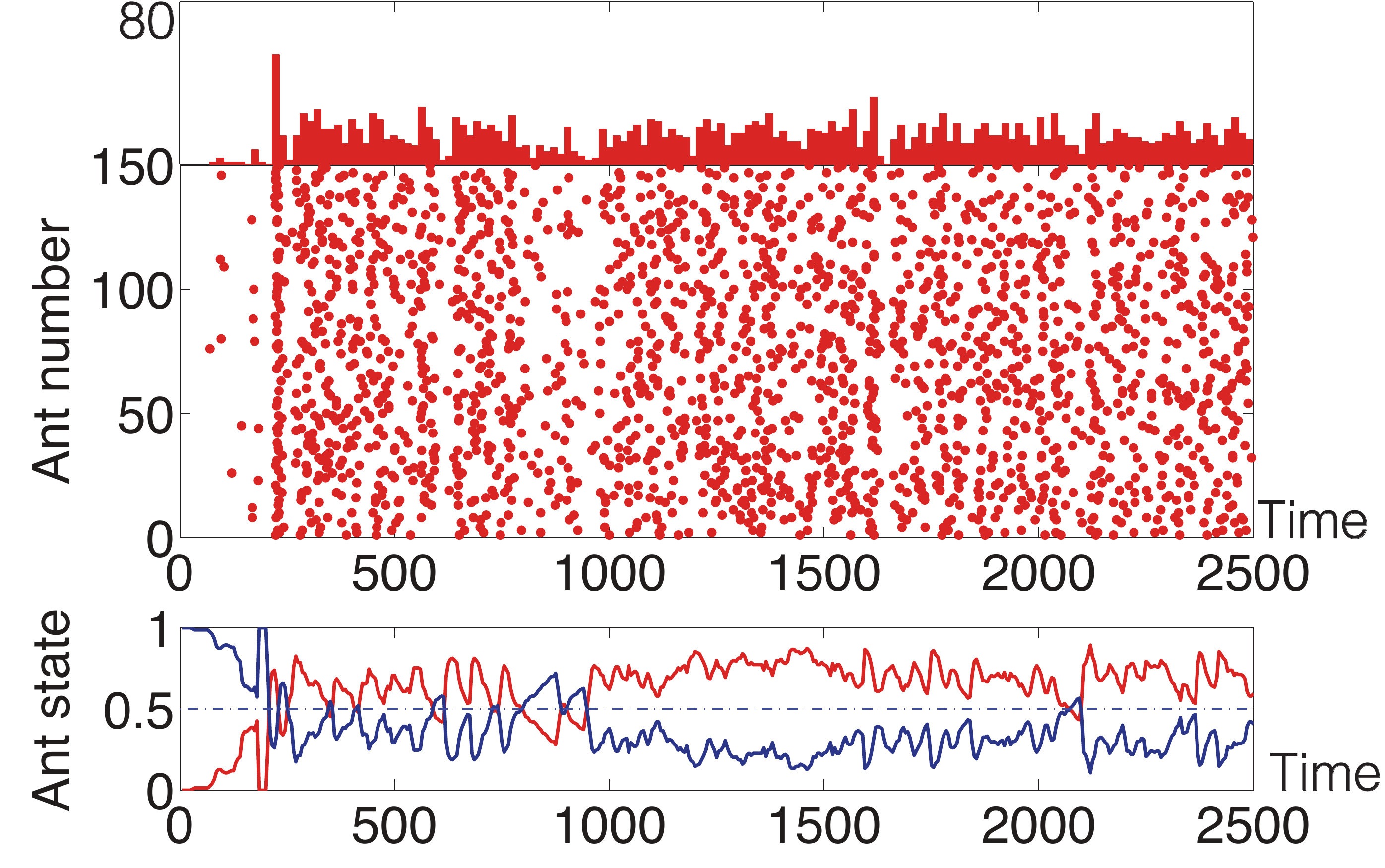}
\includegraphics[width=0.49\textwidth]{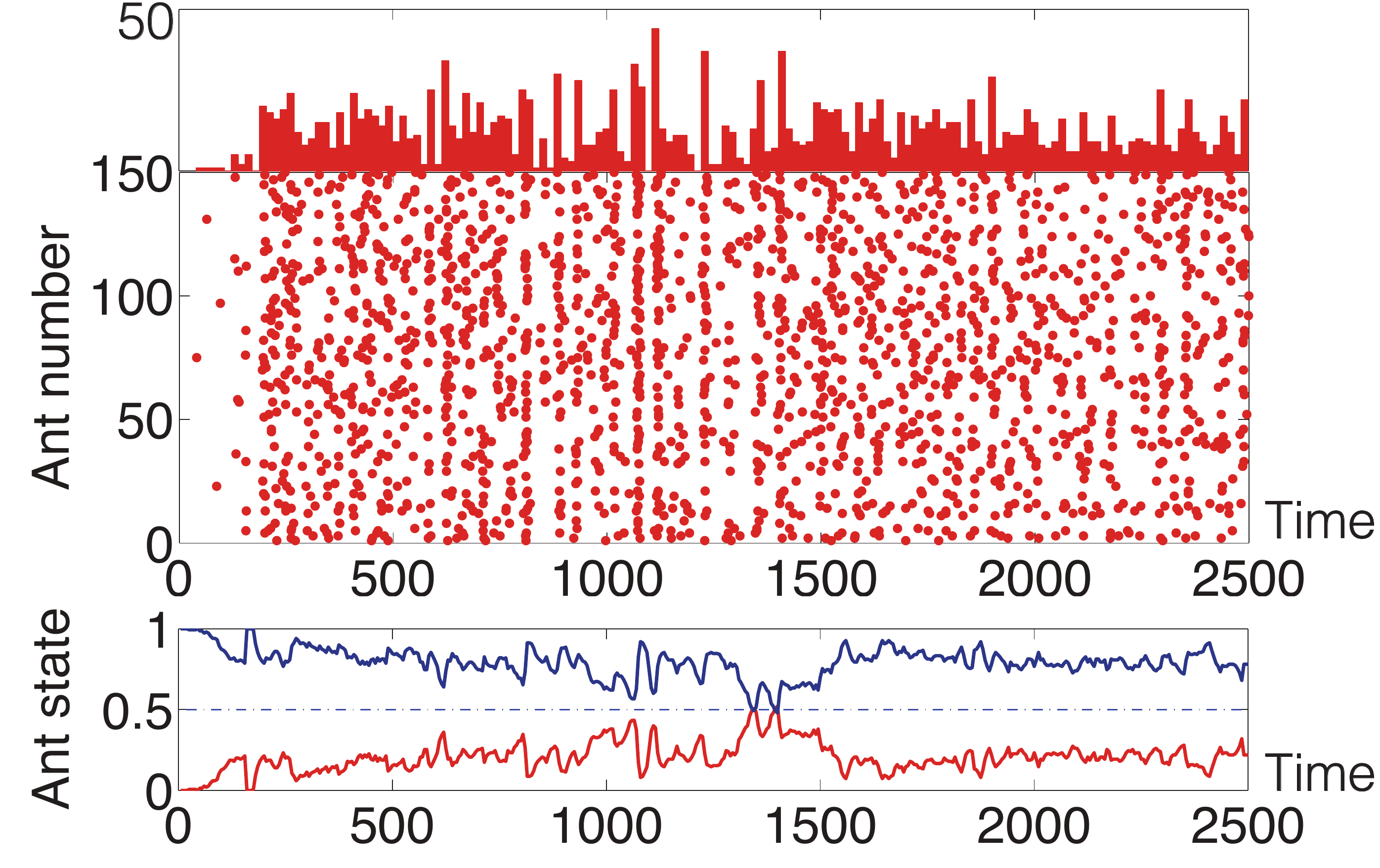}\\
\includegraphics[width=0.48\textwidth]{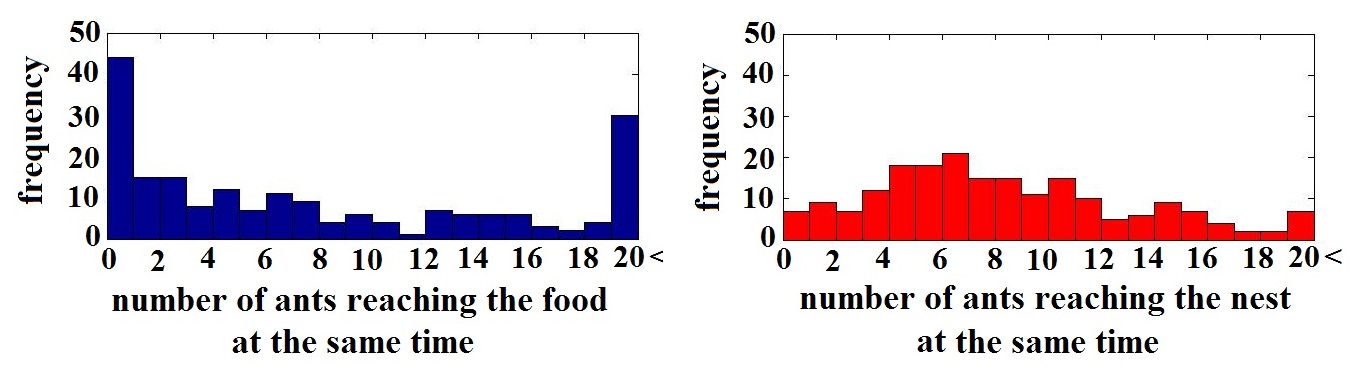}
\includegraphics[width=0.48\textwidth]{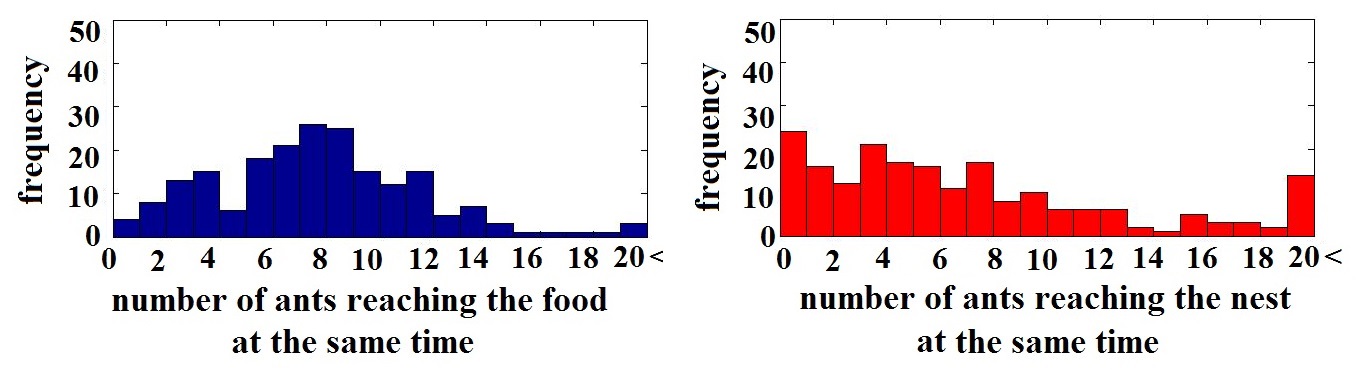}

\caption{Synchronization in the system depends on the diffusion properties of the pheromones.
Once ants find the food source for the first time they start a repeated pattern of visiting the food source and the nest site alternately. The times of the visits depend on diffusion constants $D_A$, $D_B$ of the two pheromones $A$, $B$ present in the system. The simulation results with small $D_A$ ($D_A=1$ mm$^2$/s, $D_B=5$ mm$^2$/s) are given on the left and the simulation results with small $D_B$ ($D_A=5$ mm$^2$/s, $D_B=1$ mm$^2$/s) are on the right. While for small $D_A$ the times of food discovery show synchronisation, this situation is reversed for small $D_B$ where the times of nest discovery show increased synchronisation. The top two rows of panels show the times of events at which each ants arrives at the food source and the nest (food in blue and nest in red) in addition to histograms in which arrivals are binned into 150 equally spaced intervals. In the third row we show the proportion of ants in each phase. This signal oscillates mildly on a small time-scale which aligns with the synchronized change of state seen in the histograms of the top two rows. The bottom row of panels show histograms of the frequency of events  (10 second windows of time) in which a given number of ants find the food source (blue) or the nest (red) for the two different ratios of diffusion coefficients. 
\label{fig6}
}
\end{figure}

By investigating the behaviour of individual ants in the model we discovered some other interesting and unexpected behaviours. Surprisingly, the model ants demonstrate synchronisation in their discovery of the nest or the food source when given a single fixed non-depleting food source. We evidence this in Figure~\ref{fig6} where we show (top two rows) plots of the times at which each of the ants finds the food (blue dots) or the nest site (red dots) for situations in which the diffusion coefficient of pheromone A is greater than that of pheromone B (left column) and vice versa (right column). When a significant proportion of ants find the food or the nest nearly-simultaneously we observe a nearly-continuous vertical line in the raster plot and a correspondingly high peak in the histogram for absolute arrival time.

Interestingly, there is an asymmetry between the events of the food and the nest discovery that depends on the absolute magnitudes of the diffusion coefficients of the two pheromones. When $D_A$ is small the ants have a tendency to  discover the food synchronously. They also show some synchrony in finding the nest but the effect is less pronounced. On the other hand, for $D_B$ small nest discovery events appeared to be synchronous. It is important to note that after some sufficiently long transition time the food and nest (if of the same size) can be only distinguished by the diffusion constants of the associated pheromones.

Synchrony in ant food/nest discovery should mean that, within a given short (but fixed) time interval, either a large number of ants or very few ants find the food/nest. On the contrary, if ants are distributed and travel uniformly between the nest and the food source (and vice versa) then we should see approximately the same number of discovery events within each fixed time interval. We measured the number of food/nest finding events within contiguous 10-second time windows and plotted their frequencies as histograms in the bottom row of Figure~\ref{fig6}. The observed bimodal distribution for the food-finding event sizes and more uniformly spread event sizes for nest-finding events when $D_A$ is small and $D_B$ is large (and vice versa for $D_B$ small and $D_A$ large) confirms a non-symmetric synchrony pattern. We found similar results for each of the repeat simulations we investigated.

\begin{figure}[H]
	\centering
	\includegraphics[width=\textwidth]{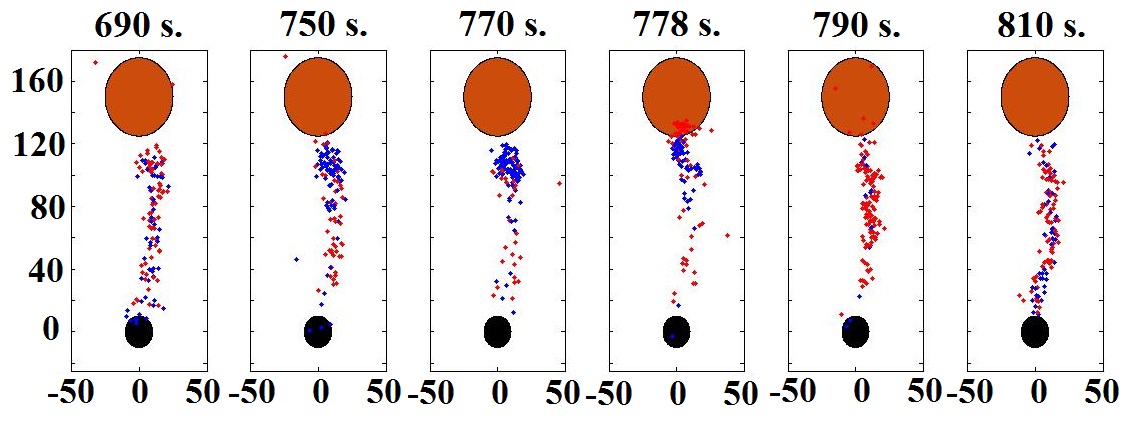}
	\caption{The distribution of ants along the trail varies in time. We represent food-searching ants by blue dots and nest-searching ants by red dots. The first panel shows an approximately uniform distribution of ants along the trail. At  $t=750$ (panel 2) there is a cluster of food-searching ants milling with few nest-searching ants close to the food source. The milling cluster becomes larger and more apparent by the $t=770$ (panel 3) when the majority of ants following the trail are clustered. We observe a synchronized event of food finding around $t=778$ (panel 4). When the mill is broken, ants find the food source synchronously and change their state from food-searching to nest-searching. After the synchronised event, at $t=790$ (panel 5) there are more nest-searching than food-searching ants on the trail but these are approximately uniformly distributed along it. At $t=810$ (panel 6) ants are again uniformly distributed in both states along the whole trail.
} 
	\label{fig7}
\end{figure}

We can gain insight to these synchronization events by considering the ants' positional distribution along the trail before, during and after a synchronized food/nest-finding event. We present such images in Figure~\ref{fig7}. At some point before a synchronized food/nest-finding event we observe an approximately uniform distribution of food-searching and nest-searching ants along the trail (panel 1). However, in the case of small $D_A$, at some later time (panels 2 and 3) food-searching ants become clustered close to the food source (and vice versa close to the nest for small $D_B$). These clustered ants exhibit a milling pattern. This behaviour immediately precedes the synchronized food-finding event, that occurs when the milling cluster is broken (panel 4). Afterwards many ants simultaneous change state and disperse so that there are more ants in the nest-searching state uniformly distributed along the trail (panels 5 and 6). After some time the ratio of ants in both states and their distribution along the trail uniformly stabilizes until the next synchronized event appears. 

In order to understand why the food finding events may be synchronised while the nest finding events are not (if $D_A$ is small and $D_B$ is large) we need to consider the pheromone concentrations. $D_A<D_B$ implies that the concentration profile of pheromone A will be steep in comparison to the gradient of pheromone B.
In particular ants searching for the food source, will experience a shallow concentration of pheromone B around the food location. This shallow gradient provides less information about the direction of the food source than a sharp gradient. Ants may be further obfuscated by the local pheromone peaks of the ants making their return to the nest, marking the position of the food. So instead of marching towards food the food-searching ants may follow these recent nest-searching ants temporarily. Equally some of the local nest-searching ants may also follow the food-searching ants, together forming a local mill. This mill may be broken by the effects of randomness or decay of the local pheromone peaks after which the ants, that are now clustered in space close to the food source, find the food synchronously. The likelihood of forming such a mill around the nest, where the gradients are large, is much smaller. Nevertheless, we may also observe a few synchronisation events around the nest.

\section{Discussion} 

We proposed a mathematical model of ant foraging involving two pheromones. The aim of the model was to capture the most important aspects of foraging behaviour of ants whilst making the model simple enough to understand and interrogate. As such, we include just two pheromones which we have shown to be sufficient to achieve a  variety of realistic biological behaviours.  The motion of individual ants is described as a correlated random walk with a deterministic contribution which allows ants to climb up gradient of pheromone (corresponding to the resource they are currently seeking) and a stochastic contribution which diminishes in strength when the pheromone signal is very strong. Pheromones both diffuse and decay and mark either the location of a food-source or a nest. By prohibiting direct interaction between ants, we ask whether interactions through a global but dynamic pheromone field are sufficient to allow realistic foraging strategies.

We investigated our mathematical model numerically, under various parameter regimes and environmental conditions. We demonstrated that our model ant colony is able to form, follow and maintain a narrow trail between the nest and the discovered food source. Furthermore the system is able to adapt to dynamic alterations in the environment such as changes in strength and placement of food sources.    

A particularly important part of the model is the randomness in the motion of each ant. The stochastic behaviour allows ants to explore the environment for food and, once found, to refine the trail between the nest and the food source. I also endows the colony with adaptability since ants are able to deviate from established trails and discover new food sources.

Foraging efficiency also depends on the ability to exploit multiple food sources with a limited group of ants. There is clearly a trade-off between having a strong pheromone trace that is easy to follow but restricts ants' abilities to explore their surroundings and a weaker trail which allows for a more through exploration of the environment at the cost of a loss of reliability in locating known food sources. However, in our model, we have found that the locations of the food sources also influence the number of food sources that a colony can exploit simultaneously. We found that maintaining two trails with a given limited number of ants is less likely if the food sources are close to each other (the two nests located on a circle have small angular distance with the maximum distance achieved when the food sources are on diametrically opposite sides of the nest. This is because the trails of two food sources placed at small angular distance interfere with each other and ant face a choice between the two. As 
soon as the symmetry between the pheromone intensities of the two trails is broken, positive feedback reinforces the stronger of the two trails trails. Random exploration may lead ants to reinvent the forgotten trail later, however, we found that the probability of forming just a single trail decreased as the subtending angle between the food sources increased. This phenomenon can be thought of as a stochastic bifurcation between the states in which one or two trails can be maintained, with the subtending angle as the bifurcation parameter. It would be interesting to further generalise the investigation of this phenomenon by adding more food sources. This may be a particularly important consideration in a study of ant colonies in which the population changes dynamically depending on food availability, but we leave the study of these more complicated scenarios for a future work.

We have observed some unexpected phenomena when studying the behaviour of ants within a well-established trail. The trajectories of ants follow the patterns have been observed in the real ant colonies \cite{couzin, dene2} -- oscillating motion of individuals along the trail and milling behaviour. More surprisingly, the ants show synchronisation as they switch between the food-finding and nest-finding state. This synchronisation is not symmetric, i.e. they synchronously find the food but consequently desynchronise on the return to the nest or vice versa. This  asymmetry between the states is depends on the ratio of the diffusion coefficients of the pheromones, where the synchronization event occurs at the resource that marked by the pheromone with the larger diffusion coefficient. 
 
We have used a simple model with only two attracting pheromones that lead ants from nest to the food source and back. Since ants in our model have no memory, and do not use sight or other intrinsic environmental cues, at least two pheromones are required. A single guiding pheromone would allow ants to distinguish between the signal leading to the nest and that leading to the food source. If ants are programmed to follow a single pheromone, positive feedback by ant simultaneously following and depositing this pheromone could create an artificial high pheromone gradient and cause ants to mill futilely until all the pheromone deposited had diffused/decayed. It is also possible that we could extend the model by allowing ants to employ more that two pheromones potentially allowing the ants to achieve more complex tasks. However, we have shown that ants can carry out basic foraging behaviour using just two.

In this paper we have employed a simple model in order to replicate the foraging behaviours of ants. We have demonstrated that our model ants are capable of achieving a number of complex tasks such as foraging with fixed and variable food sources. There are many tasks which real ants have been shown to be capable of with which we have yet to test our model ants \cite{reid,rob,nico}. For example, we have not explored whether our model ants are able to discern different quality food sources or nests (perhaps by differential pheromone deposition) or what choices are made when food sources are placed at variable distances from the nest. These are questions of significant biological interest which we hope to address in future publications.

\linespread{1.1}
\begin{table}[H]
\begin{center}
\begin{tabular}{|c|c|c|p{7cm}|}
\hline
 PARAMETER & VALUE & REFERENCE & EXPLANATION \\ \hline
\multicolumn{4}{|c|}{\textbf{ANT}} \\ \hline
body length& $2.1$&  2 - 15 mm \cite{bicak} & body length of an ant (step size is defined as two times the body length) \\ \hline
velocity&$ 8.4 $& 5 - 13 mm/s \cite{bicak}& velocity of an ant is constant \\ \hline
antennae angle& $\pm\frac{\pi}{6}$& $\pm\frac{\pi}{6}$ \cite{couzin}& angle in which antennae differ from straight direction\\ \hline
antennae length&0.7& $\frac{1}{3}$ body length& length of the antenna needed to set the exact point of pheromone perception \\ \hline
\multicolumn{4}{|c|}{\textbf{MOTION}} \\ \hline
$\alpha$&$10^{-5}$& \cite{perna}& threshold from Webber's law \eqref{Weber} \\ \hline
$\sigma$&$1.0991$& data from \cite{bicak}& standard deviation of random directional change \\ \hline
\multicolumn{4}{|c|}{\textbf{PHEROMONES}} \\ \hline
$ D_A, D_B$ &  1 , 5 & $1$ $ mm^2/s$ \cite{ever}& diffusion constants\\ \hline
 $\delta_A, \delta_B$ &  200, 100  & s & decay of pheromones \\ \hline
  m & 0.01 & g &amount of pheromone deposited in one deposition \\ \hline
 $c_\text{tresh}$ & $10^{-11}$ & $10^{-11}$ g \cite{ever} & pheromone minimum detectable  \\ \hline
$c_{\min}$ & $10^{-5}$ &g & decrease of randomness in equation \eqref{sigma} \\ \hline
$t_A, t_B$& 80 , 80  & s  &for how long a pheromone is deposited \\ \hline
\multicolumn{4}{|c|}{\textbf{SIMULATION NUMERICS}} \\ \hline
ants number (N)& 150 & - & number of ants in the system \\ \hline
$[x_\text{nest}, y_\text{nest}]$&[0,0]&-& position of the nest\\ \hline
$[x_\text{food}, y_\text{food}]$&[0, 150]&-& position of the food\\ \hline
h& 0.5 & - & step size on the numeric grid\\ \hline
multi& 0.1 & - & multiplicator in the numeric computation of pheromone field \\ \hline

\end{tabular}
\end{center}
\caption{Parameters of the model with references and their settings for the basic simulation.}
\label{tab1}
\end{table}

\normalsize

%

%

%
%
%
%


\end{document}